\documentclass[a4paper,twocolumn,10pt]{article}
\usepackage{eurosis}
\usepackage{url}
\usepackage{natbib}
\usepackage{latexsym}
\usepackage{caption}
\usepackage{subcaption}
\usepackage{graphicx}
\usepackage{graphics}
\usepackage{adjustbox}
\usepackage{mathptmx}
\usepackage[T1]{fontenc}
\usepackage{xcolor}
\usepackage{wrapfig}
\usepackage{float}
\usepackage{amsmath}
\usepackage{xspace}
\usepackage{authblk}
\usepackage{hyperref}
\begin{document}

\title{A MULTI-OBJECTIVE COMBINATORIAL OPTIMISATION FRAMEWORK FOR LARGE SCALE HIERARCHICAL POPULATION SYNTHESIS}
\author{%
  \begin{minipage}{0.5\textwidth}
    \centering
    Imran Mahmood, Nicholas Bishop,
    
    Anisoara Calinescu, Michael Wooldridge\\
    Department of Computer Science\\
    University of Oxford\\
    Wolfson Building, Parks Road, Oxford, OX1 3QD, UK
  \end{minipage}%
  \begin{minipage}{0.5\textwidth}
    \centering
    Ioannis Zachos\\
    Department of Engineering,\\
    University of Cambridge\\
    Trumpington Street, Cambridge CB2 1PZ, UK
  \end{minipage}
}
\date{}
\maketitle
\thispagestyle{empty}

\keywords{Agent-based simulations, hierarchical population synthesis, multi objective combinatorial optimisation, genetic algorithms.}

\begin{abstract}
In agent-based simulations, synthetic populations of agents are commonly used to represent the structure, behaviour, and interactions of individuals. However, generating a synthetic population that accurately reflects real population statistics is a challenging task, particularly when performed at scale. In this paper, we propose a multi objective combinatorial optimisation technique for large scale population synthesis. We demonstrate the effectiveness of our approach by generating a synthetic population for selected regions and validating it on contingency tables from real population data. Our approach supports complex hierarchical structures between individuals and households, is scalable to large populations and achieves minimal contigency table reconstruction error. Hence, it provides a useful tool for policymakers and researchers for simulating the dynamics of complex populations.
\end{abstract}

\section{INTRODUCTION}

Population synthesis plays a crucial role in generating meaningful emergence structure from agent-based simulations. Common applications include urban planning, transportation and public health modelling \cite{smith2017population}. A Synthetic Population (SP) is a simulated population that matches key demographic, social, economic, and geographic characteristics of a real world population. SPs assimilate real world data, which are often limited, sensitive, unavailable, or costly to obtain \cite{barthelemy2013synthetic,horl2021synthetic} for modelling, and policy scenario testing. They are integral for initialising agent-based simulations (ABS), due to their realism, privacy preservation, flexibility\footnote{A flexible synthetic population algorithm can adjust its parameters and modelling assumptions to account for a variety of available data and different modelling goals and objectives}, and reproducibility \cite{ye2009methodology},. ABS is a novel paradigm that helps the study of complex adaptive systems through a systematic bottom-up abstraction of the system, where the behaviour of individual agents and their interactions are studied to understand and predict the dynamics of these complex systems \cite{macal2005tutorial,bonabeau2002agent,mahmood2022facs}. They are used to explore and evaluate different assumptions, interventions, or policies \cite{wu2022synthetic}. 
This study aims to: (1) develop a methodology for generating synthetic populations at a selected scale and region that accurately matches the aggregate demographic characteristics and respects their hierarchical structure of the target population; (2)demonstrate the flexibility of the proposed approach in addressing diverse simulation requirements; (3) evaluate the synthesised population in terms of accuracy and computational efficiency. To achieve these objectives, we offer multi-objective combinatorial optimisation using the Non-dominated Sorting Genetic Algorithm II (NSGA-II) \cite{deb2002fast}. NSGA-II combines genetic algorithms with non-dominated sorting to efficiently search for optimal solutions in a combinatorial space. The optimisation objectives consist of individual demographic and spatial distributions, while allowing for weighting of these objectives depending on the simulation context. Key contributions of this paper are listed as follows:
\begin{itemize}
    \item The development of a novel methodology for generating country-scale synthetic populations that accurately represent the demographic structure, using multi-objective combinatorial optimisation techniques.
    \item The assessment of the representativeness and  accuracy of the generated population, the scalability of the approach, and the computational efficiency of the generation process.
    \item Presentation of a case study demonstrating the generation of synthetic population of the selected regions in the city of Oxford, providing insights into the practical implementation of the proposed methodology.
    \item Discussion of the advantages of our approach and of future work directions in the field of synthetic population generation for ABS.
\end{itemize}

Different approaches have been used for synthetic population generation. The approaches are grouped into three categories: Synthetic Reconstruction (SR), Combinatorial Optimisation (CO), and Statistical Learning (SL). SR methods: \cite{jiang2022method}, \cite{fabrice2021comparing}, \cite{ponge2021generating}, \cite{pritchard2012advances}, \cite{muller2017generalized} involve fitting and allocation to generate synthetic populations by adjusting weights and cell counts. CO methods: \cite{chapuis2022generation}, \cite{harland2012creating,wu2022synthetic}, \cite{kurban2011beginner}, \cite{chen2016genetic},\cite{srinivasan2008procedure} involves finding the best solution from a set of possibilities using optimisation techniques. SL methods: \cite{sun2018hierarchical}, \cite{farooq2013simulation}, \cite{saadi2016hidden} focus on the joint distribution of attributes and uses machine learning and probabilistic methods. Each approach has its strengths and considerations regarding accuracy, computational requirements, and data availability. SR methods simplify assumptions for accurate results with high-quality marginal data, while SL techniques capture complex attribute relationships but may be computationally demanding and require extensive training data. In contrast, CO offers a flexible approach, optimising multiple objectives, especially with a hierarchical structure, and can handle data sparsity based on problem nature and data quality. However, CO may need significant computational resources and tuning. The approach choice depends on goals, constraints, data availability, and resources. Our proposed CO-based approach efficiently generates a customizable representative large-scale population by utilising multi-objective optimisation to fit individual attributes with census data and a hierarchical structure.

\section{Proposed Approach}
In this section we discuss our proposed approach. First, we describe how synthetic population generation may be formulated as a multi-objective optimisation problem. Next, we discuss how the NGSAII genetic algorithm as a multi-objective evolutionary optimisation method to generate and optimise synthetic populations with respect to census contingency tables. As a proof of concept, we conduct a case study of the Oxfordshire region, using the UK census data to generate a hierarchical population of persons and households. At the end we discuss the results and we evaluate the proposed approach. 

\subsection{Problem Formulation}

\textbf{Given:}
A set of selected attributes in a real population where each attribute $A_i$ has a set of categories (or groups, e.g., age $= 0-5, 6-10,\dots 81-85$), with respective frequencies $F_{A_i}$ in the real population:
\begin{equation}
     A = \{a_1, a_2, \dots, a_n\}
\end{equation} 
\begin{equation}
     C_{A_i} = \{c_{A_i,1}, c_{A_i,2}, \dots, c_{A_i,m_i}\}
\end{equation}
\begin{equation}
     F_{A_i} = \{f_{A_i,1}, f_{A_i,2}, \dots, f_{A_i,m_i}\}
\end{equation}

\textbf{Objective:} Generate a synthetic population that closely resembles the real population's distribution of each attribute. Let $X$ be an individual \footnote{Here we refer to an individual as a candidate solution, not a person in the synthetic population} in the population, representing a synthetic population. For each attribute $A_i, i = 1, 2, \dots, N$, the objective function is defined as: \textit{Minimise}
\begin{equation}
     O_i(X) = \sum_{j=1}^{m_i} \lvert f_{A_i,j} - f'_{A_i,j}(X)\lvert,
\end{equation}
where $f'_{A_i,j}(X)$ is the frequency of category $c_{A_i,j}$ in the synthetic population $X$. 

\textbf{Goal:} To find a synthetic population $X^*$ that minimises all objective functions $O_i(X)$, for $i = 1, 2, \dots, N$.

Generating a synthetic population involves creating a sample of individuals and households with specific characteristics that closely resemble the actual population. The goal is to capture the distributions of selected attributes found in the real population while preserving privacy. Contingency tables, which display the relationship between categorical variables, are used to describe statistical relationships between population characteristics. These tables help analyse patterns and trends among different demographic groups. By recreating the frequency distributions given by contingency tables, the representativeness of a synthetic population can be measured. Trade-offs may be necessary when fitting multiple contingency tables, as some may be more important than others depending on the application. Our approach allows practitioners to naturally balance objectives and obtain a synthetic population that suits their needs. In this study, we validate our methodology using cross tables from the 2011 UK census \cite{nomis}, which include bivariate and trivariate tables that combine different attributes. In this paper, we have considered trivariate contingency tables (e.g., Sex:Age:Ethnicity, Sex:Age:Religion, Sex:Age:Qualification -- see Figure \ref{fig: tables}).

A multi-objective optimisation algorithm can optimise two or more objectives simultaneously. This algorithm generates a set of Pareto-optimal solutions, providing a balance between objectives. The algorithm iteratively evolves a population of candidate solutions by applying genetic operators (selection, crossover, and mutation) while considering all the objectives having different weights according to their significance (e.g., in certain use-cases the economic attributes of persons and households may be more significant than ethnicity and religion). The Pareto-optimal solutions represent trade-offs between different distributions, allowing decision-makers to choose the most suitable synthetic population based on their requirements. Motivated by these features, as well as the complex hierarchical structure of synthetic populations, we employ the NSGA-II algorithm. 

\subsection{Multi-Objective Combinatorial Optimisation using Genetic Algorithms}
We formulate population and household synthesis as a multi-objective combinatorial optimisation problem. We first present a brief primer on genetic algorithms (GAs) \cite{wirsansky2020hands}, providing rationale for our use of this approach in population synthesis. GA is a type of evolutionary computation technique inspired by the process of natural selection. Genetic algorithms maintain a population of candidate solutions, which reproduce over multiple generations. In the context of our work, candidate solutions correspond to synthetic populations.  A predefined selection process is used to determine which candidate solutions may reproduce at the end of each generation. The success of a candidate within the selection process is determined by their fitness, which is evaluated via a fitness function. The fitness of a synthetic population describes how well it recreates the frequency distributions of contingency tables. Reproduction consists of both crossover and mutation. We provide more details regarding each component of a genetic algorithm below:
\begin{itemize}
    \item \textbf{Selection:} This is the process of choosing individuals from the current population based on their fitness values. Selection favours individuals with higher fitness values (or lower in our case of minimisation the error), ensuring that the best solutions have a higher probability of being chosen for reproduction. Common selection methods include tournament selection, roulette wheel selection, and rank-based selection \cite{deb2011multi}.

    \item \textbf{Crossover (or recombination):} This operation combines the genetic material of two parent individuals to produce one or more offspring. The goal of crossover is to create new individuals that inherit the best traits from their parents, potentially leading to better solutions in the next generation. There are various types of crossover operators, such as one-point crossover, two-point crossover, and uniform crossover.
    
    \item \textbf{Mutation:} This operation introduces small random changes in an individual's genetic material. Mutation helps maintain diversity in the population and prevents premature convergence to sub-optimal solutions. Mutation operators can vary depending on the problem representation; for example, bit-flipping mutation for binary strings or Gaussian mutation for real-valued representations.

    \item \textbf{Fitness Evaluation:}The fitness function evaluates the quality of each individual in the population based on how well they solve the given problem. It assigns a fitness value to each individual, which is then used for selecting and determining the best solutions. The fitness function is problem-specific and designed to guide the search towards optimal or near-optimal solutions.
\end{itemize}

\begin{figure}[h]
    \centering
    \includegraphics[width=0.7\columnwidth]{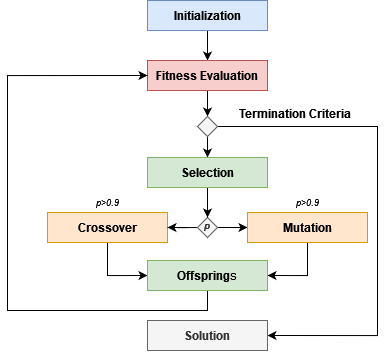}
    \caption{Genetic Algorithm Flow\label{fig: flow}}
\end{figure}

Figure \ref{fig: flow} shows the flow of the genetic algorithm. 

The Non-dominated Sorting Genetic Algorithm II (NSGA-II) is a popular multi-objective optimisation algorithm that extends the conventional GA framework described above to handle problems with multiple conflicting objectives. NSGA-II employs a fast non-dominated sorting approach to categorise the individuals into different levels of Pareto frontier. Moreover, NSGA-II  employs a crowding distance metric to maintain diversity in the population, preventing premature convergence to sub-optimal solutions. Using NSGA-II for synthetic population generation offers several advantages over traditional optimisation techniques: (i) it can optimise multiple aspects of the population simultaneously without requiring objectives to be combined into a single value; (ii) It employs a Pareto-based approach to identify non-dominated solutions that represent the best trade-offs between objectives, allowing stakeholders to choose the most suitable synthetic population; (iii) It preserves diversity in the population by using a crowding distance metric and incorporates elitism to preserve the best solutions found in previous generations; and (iv) It is scalable to handle problems with a large number of objectives or decision variables.

\begin{figure*}[h]
    {
    \centering
    \includegraphics[width=0.7\textwidth]{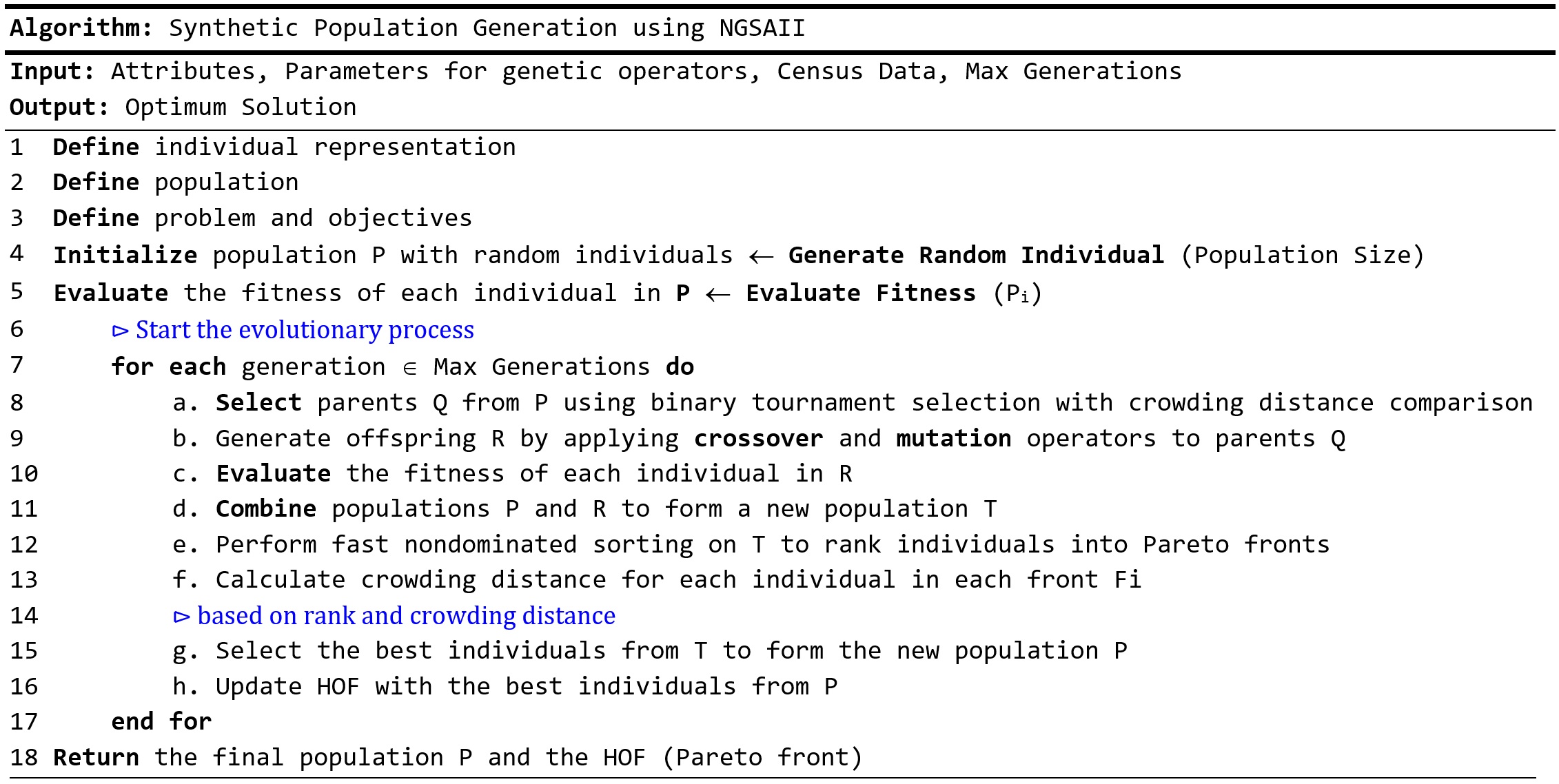}
    \caption{Synthetic Population Generation using NGSAII \label{fig: algorithm1}}
    }
\end{figure*}

\begin{figure}[h]
    {
    \centering
    \includegraphics[width=\columnwidth]{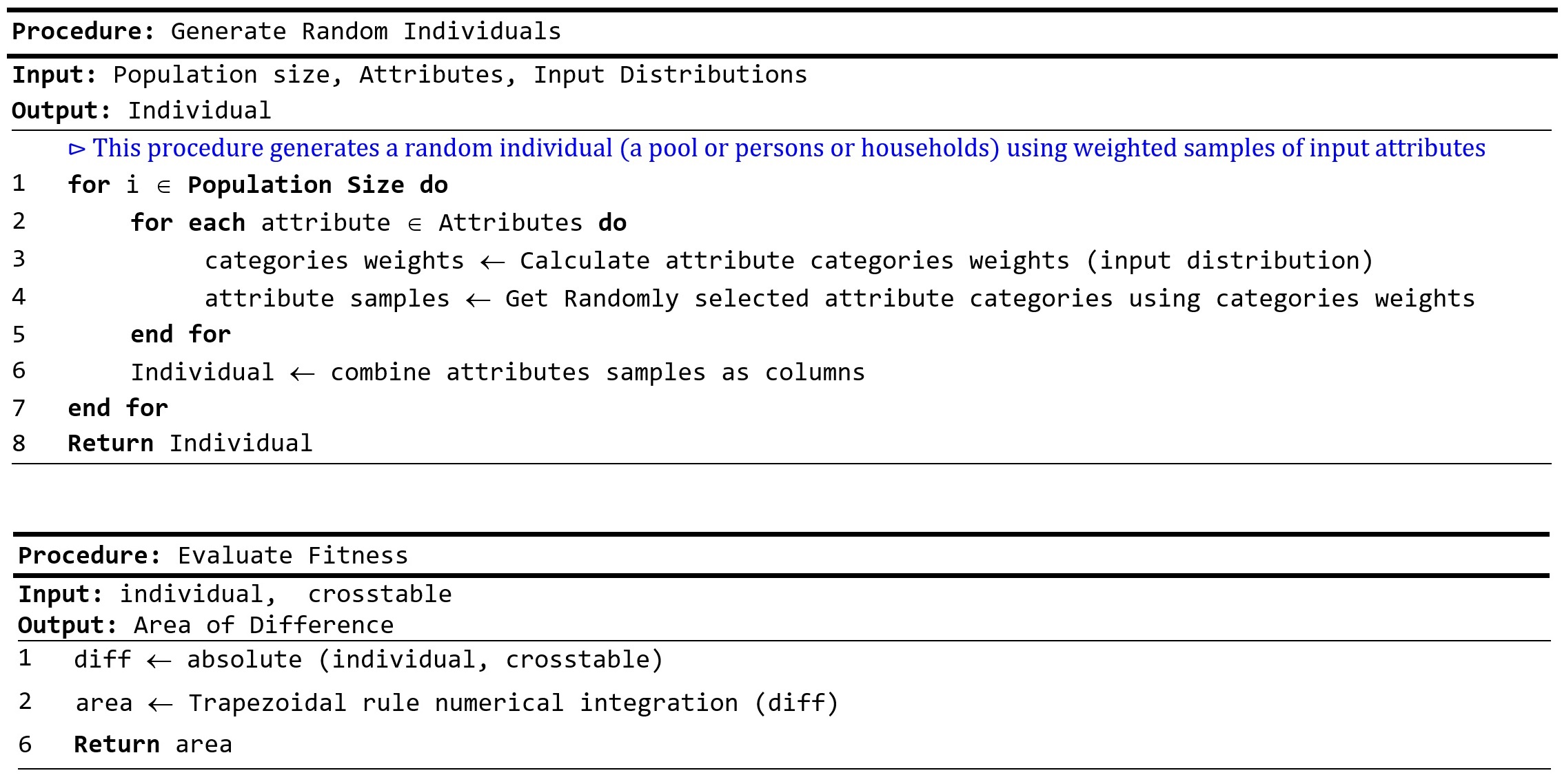}
    \caption{Individual generation and Fitness Calculation
    \label{fig: algorithm2}}
    }
\end{figure}

Our proposed algorithm is given in Figure \ref{fig: algorithm1}. At first we define the data structure and encoding of the individuals (solutions) and the population, which will store a group of individuals. The problem is to generate random samples of persons and households and then allocate persons into households using census data constraints. The objective is to minimise the difference of error between the generated samples and the actual census data, for each selected attribute,therefore the problem is multi-objective (lines 1--3). Then we create an initial population by generating random individuals \footnote{An individual is a term typically used in GA for the entity being generated. In our case it could be a person or a household} using the procedure shown in figure \ref{fig: algorithm2} (line 4). In this procedure first we calculate attribute weights from the census data tables. Then we generate random samples for each attribute and finally combine these attribute samples to form a set of individuals. We use a rule-based validation routine to accept or reject a random combination if it does not satisfy certain rule (e.g., an individual of age $< 18$ cannot be married). Next we calculate the fitness value of each individual in the initial population using the fitness evaluation function using the procedure shown in Figure \ref{fig: algorithm2}. This is a proposed method of calculating the total area of the difference between the two curves of the generated sample and the actual data, using the Trapezoidal numerical integration \cite{yeh2002using} (line 5). This fitness measure is more effective than conventional approaches as it captures the overall difference between the distributions and takes into account the shape and distribution of the curves. Hence, provides a more accurate measure of how well the generated population matches the target population across the entire domain. Next we create a data structure to store the best non-dominated solutions found throughout the generations, called the Pareto frontier \footnote{In multi-objective optimisation, the Pareto frontier is a set of optimal solutions that represent the trade-offs between the conflicting objectives. It is a set of solutions where no objective can be improved without worsening at least one of the other objectives.}. Now the algorithm enters into the main loop of the evolutionary process. For a given maximum number of generations, We iterate through each generation, and choose a set of parent individuals from the current population using binary tournament selection. The selection process is based on the individuals' rank and crowding distance. Crowding distance measures the distance between a solution and its neighbouring solutions in the objective space. Then create a new set of offspring individuals by applying genetic operators (crossover and mutation) to the selected parents. We use a two-point crossover method by selecting two random points along the length of the parent chromosomes and swapping the segments between the two points to create new offspring. For mutation we implemented a swapping technique which randomly selects an attribute of an individual and swaps its value with another individual. When the genetic operators are applied, we compute the fitness values for each offspring individual using the fitness evaluation function. Then we merge the current and the offspring population to create a combined population and rank the individuals in the combined population into non-dominated fronts using fast non-dominated sorting. Then we compute the crowding distance for each individual in each front, which is a measure of how crowded the solutions are in the objective space. Then we choose the best individuals from the combined population, considering both rank and crowding distance, to create the new population for the next generation. Then we add the best non-dominated solutions from the current population to the Pareto frontier  (lines 7--17). Finally we return the final population selected from the the Pareto frontier of non-dominated solutions (line 18). Selecting the best solution from a Pareto frontier  depends on the preferences. We use a weighted sum approach, by assigning weights to the objectives and select the solution with the highest weighted sum. Once the algorithm is terminated we retrieve the generated population of individuals and store it in a CSV file. 

\subsection{UK Case Study}
\begin{figure*}[h]
    {
    \centering
    \includegraphics[width=0.7\textwidth]{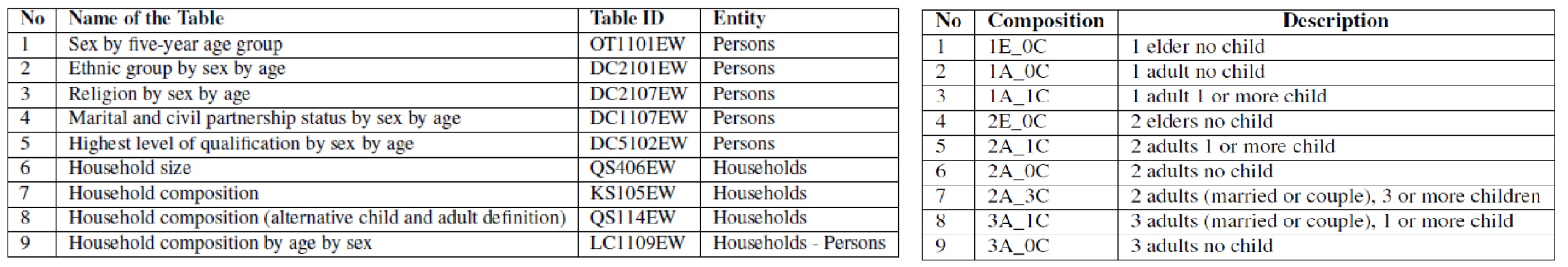}
    \caption{(a) Input Tables for fitness evaluation  (b)Household Composition Types \label{fig: tables}}
    }
\end{figure*}

This section presents the implementation details of our proposed approach In this section, we present the case study to generate a representative synthetic population of a selected region in the UK using our propose approach. Our case study is conducted at a geographical scale of Middle Super Output Areas (MSOA). There are approximately 7,200 MSOAs in England and each MSOA contains between 5,000 and 15,000 residents. They are used as geographic building blocks for analysing data and gaining insights into the distribution of characteristics across larger areas and assist in policy-making and interventions. We leverage UK Census data for the attributes of persons and households \cite{nomis}. The ethnicity in the Persons data are symbolised as: W1-W4 are categories of White; M1-M4 are mixed categories; A1-A5 are Asian categories; B1-B3 are Black and O1-O2 are Other categories. Similarly Religions are symbolised as: C=Christian, B=Buddhist, H=Hindu, J=Jewish, M=Muslim,  S=Sikh, O=Other religions, N=No religion and NS=Religion not stated. Different compositions represented in the household data are categories in Table \ref{fig: tables}.

% \begin{figure*}[h!]
%     {
%     \centering
%     \includegraphics[width=0.6\textwidth]{figures/Censusdata.png}
%     \caption{Distributions of Person and Household attributes (Census Data) \label{fig: census data}}
%     }
% \end{figure*}

\begin{figure*}[h!]
{
\centering
\includegraphics[width=0.6\textwidth]{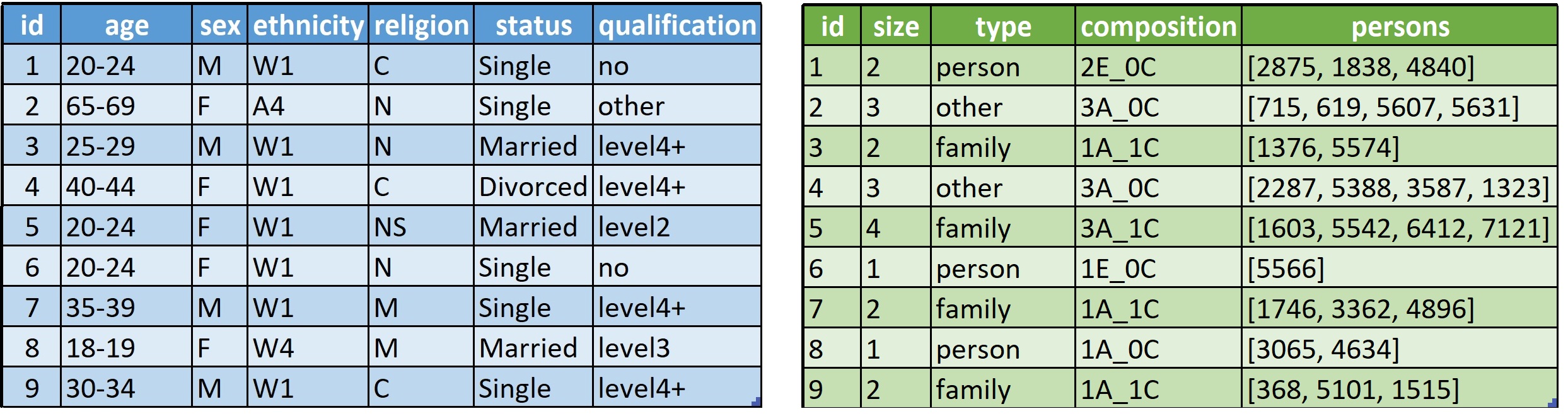}
\caption{(a) Generated Persons (b) Generated Households\label{fig: table}}
}
\end{figure*}
\begin{figure*}[h!]
{
    \centering
    \includegraphics[width=0.7\textwidth]{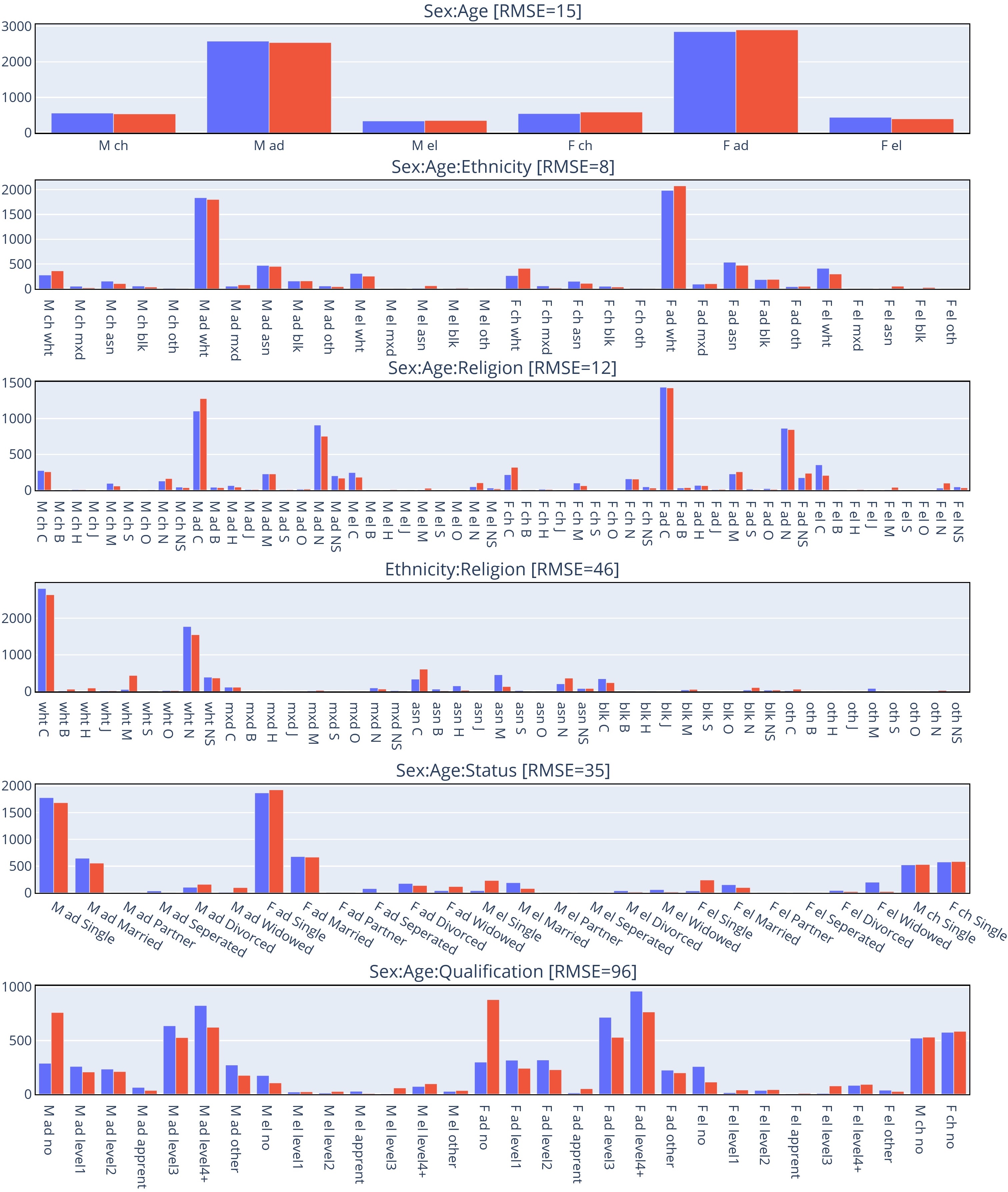}
    \caption{Generation of Persons [Blue = Actual, Red = predicted]}
    \label{fig: persons}
}
\end{figure*}

\begin{figure*}[h!]
{
\centering
\includegraphics[width=0.7\textwidth]{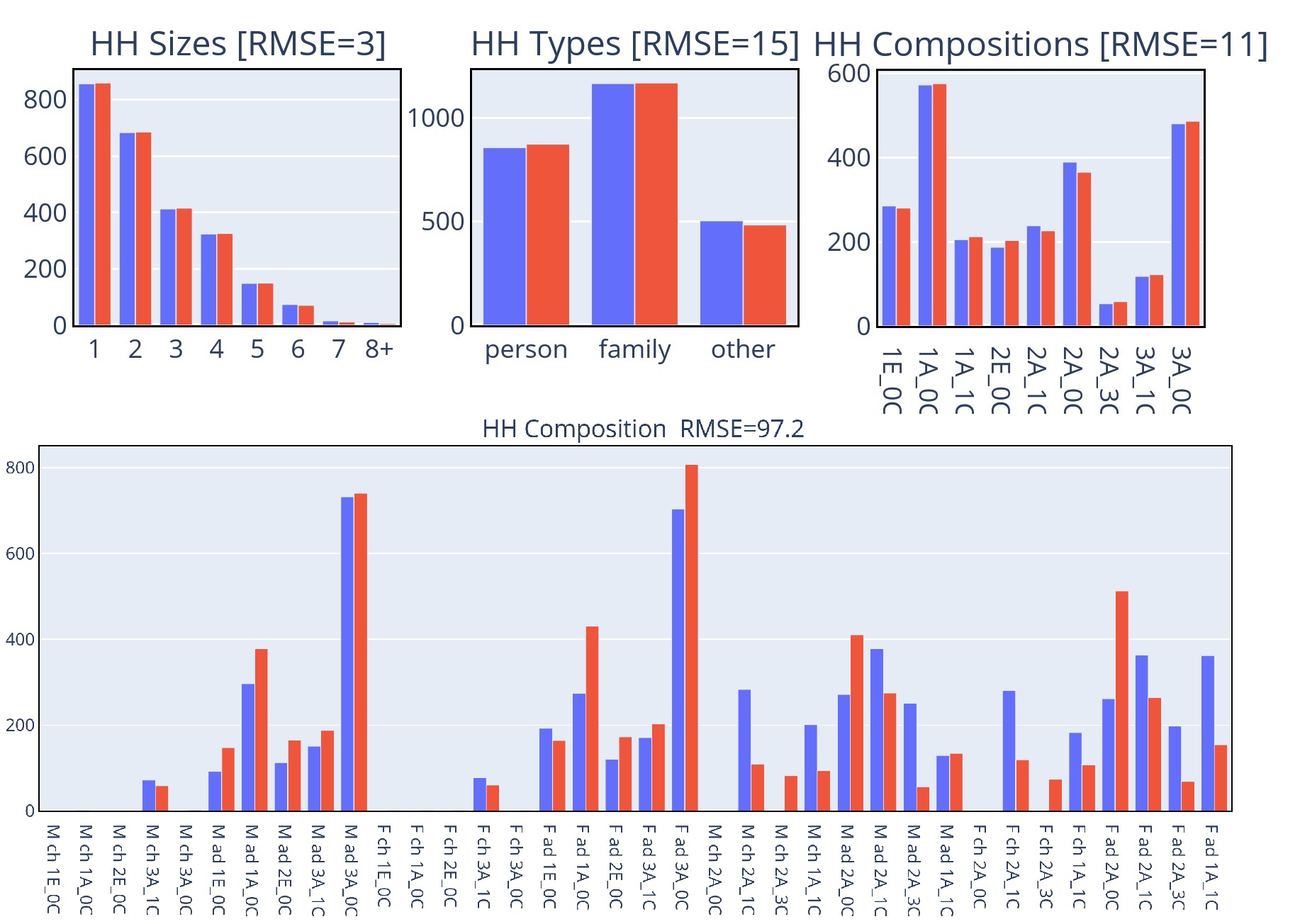}
\caption{Generation of Households and Household compositions [Blue = Actual, Red = predicted]\label{fig: households}}
}
\end{figure*}

\begin{figure*}[h!]
{
\centering
    \includegraphics[width=0.75\textwidth]{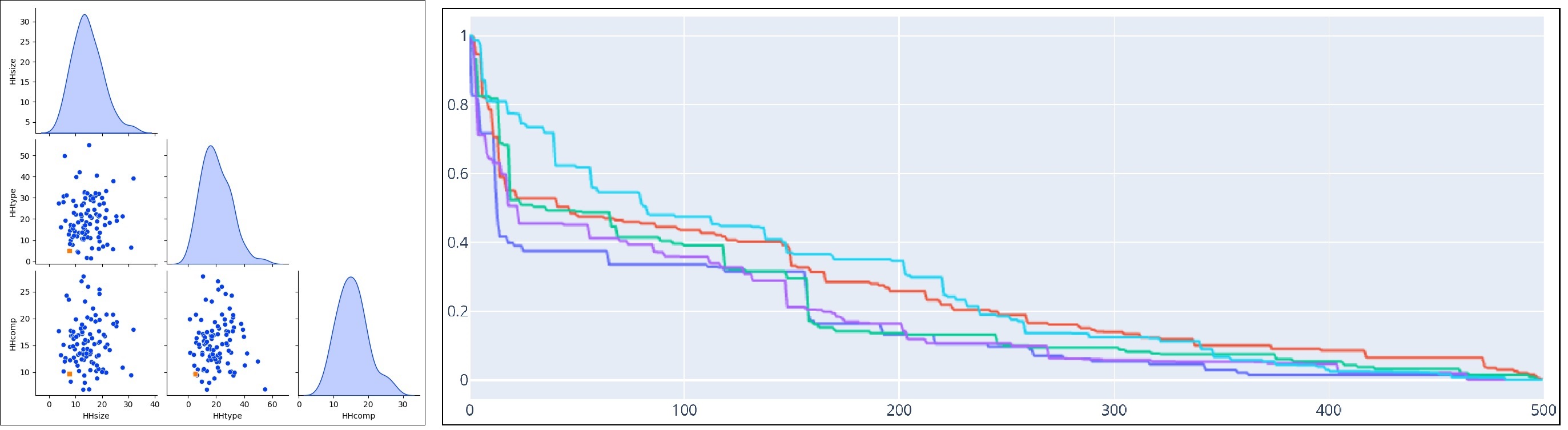}
    \caption{(a) Pareto frontier  [red = selected best solution] (b) Convergence of different objectives}
    \label{fig: pareto}
}
\end{figure*}

In this case-study we have selected two types of entities in our synthetic population: (i) Persons and (ii) Households. We aim to generate samples of persons and households according to the statistics of the selected MSOA and fit both sets using the contingency tables shown in Figure \ref{fig: tables}. Our proposed approach leverages Distributed Evolutionary Algorithms. We have extended the DEAP Python framework \cite{DEAP_JMLR2012} to support the generation of synthetic population using: (a) a variety of input data; (b) selection of individual's attributes, (c) defining multiple objectives; (d) logical design of how random individuals are generated with rule-based validation; (d) design of complex fitness evaluation criteria; (e) addition or modification of genetic operators; and finally the performance improvements using parallel processing. Our implementation is available on Github\footnote{https://github.com/imqhashmi/SynPoP-GA}.

\subsection{Results and Analysis}
This section illustrates the outputs of the execution runs and the results of our case study. We performed our analysis  on a selected MSOA at a time. It is however possible to execute multiple MSOAs in parallel in order to speed up the process of generating the entire population at the country scale. We devises the framework to operate in two stages: (i) Generating persons and (ii) Generating households, because to generate households we require persons population as input. Table \ref{fig: table} illustrates several generated samples of Persons and Households.  Figure \ref{fig: persons} shows the generation results in terms of actual and the predicted population. We group the age attributes into three categories: (a) Children (ch); (b) Adults (ad) and (c) Elders (el). Similarly we grouped ethnicity into main groups: (a) White (wht); (b) Mixed (mxd); (c) Asian (asn); (d) Black (blk) and Others (oth). The difference of sum of each generated group of attributes (red) with the actual data (blue) can be noted in the figure. We also calculated the root mean square error (RMSE) as an error measure to see the difference. In the next stage, we generated households as shown in Figure \ref{fig: households}. At this stage, we implemented the household composition by allocating individuals from the persons population into suitable households based on their attributes such as size, type and composition structure (see Figure \ref{fig: tables}). For example, in order to allocate persons in a household of size 7 and composition type: \textbf{'2A 3C'} we search and allocate two adults and 5 children from the pool of persons. Currently this allocation is not sensitive towards ethnicity, religion, or other pertinent features, and is considered as our future work. A typical run-time of a single generation for an area of 7000 persons, and population size of 100 ranges between 5-7 seconds. It takes 30-35 minutes to run 500 generations. With parallel random sampling of individuals, parallel fitness evaluation and parallel genetic operations the execution time can be substantially reduced, which is considered as our future work. After a run of 500 generations a convergence plot is generated, as shown in Figure \ref{fig: pareto}, where each line represents an optimisation objective (i.e., five objectives for each attribute of persons), X-axis shows the number of generations, and Y-axis shows the descent of normalised fitness. The rate of convergence completely depends on the genetic makeup of the feature and the operators used. When the execution is complete we generate a Pareto frontier pair plot as shown in Figure \ref{fig: pareto}. In the pair plot, each pair of objectives is placed against each other in a scatter plot, and the diagonal plots show the distribution of each objective. The plot shows the selected best solution (highlighted in red), based on our weighted sum of difference method. 

\section{Summary and Conclusion}
In this paper We present a novel approach for synthetic population creation in agent-based simulations, addressing the challenges of accuracy and representation. By employing the NGAII algorithm, a multi-objective combinatorial optimisation technique, we demonstrate the effectiveness of our approach through a case study. The results exhibit its suitability for complex and large-scale problems, offering enhanced accuracy and representation compared to traditional methods.

This case study serves as a proof of concept, validating our population synthesis approach for agent-based simulations. The focus lies in optimising multiple objectives, such as demographic characteristics, to accurately represent the target population. The findings reveal that our proposed method generates high-quality synthetic populations mirroring the target population's characteristics. Furthermore, our approach is efficient, scalable, and easily adaptable to different geographic regions, input data, and types of individuals (e.g., persons, households, cars, organisations). Notably, it excels in creating and fitting hierarchical structures using input data, enabling allocation of persons in households, assignment of cars to individuals, and allocation of workplaces to persons.

We assert that multi-objective combinatorial optimisation is a comprehensive approach for synthetic population generation, capable of simultaneously optimising multiple objectives across diverse problem domains. This work contributes to the field of agent-based modelling and simulation, opening avenues for developing more realistic and large-scale models across various domains. Future work include expanding our household composition scheme to incorporate ethnicity and religion, as well as enhancing computational efficiency through parallel processing in random individual generation, fitness evaluation, and genetic operations.

\section{Acknowledgement}
This research was supported by a UKRI AI World Leading Researcher Fellowship awarded to Wooldridge (grant EP/W002949/1). M. Wooldridge and A. Calinescu acknowledge funding from Trustworthy AI - Integrating Learning, Optimisation and Reasoning (TAILOR) (https://tailor-network.eu/), a project funded by European Union Horizon2020 research and innovation program under Grant Agreement 952215.
\bibliographystyle{eurosis}

\bibpunct[; ]{(}{)}{,}{a}{}{;}
\bibliography{refs}

\end{document}